# Phase-resolved surface plasmon scattering probed by cathodoluminescence holography

Nick J. Schilder, Harshal Agrawal, Erik C. Garnett, and Albert Polman[*]

Center for Nanophotonics, AMOLF

Science Park 104, 1098 XG Amsterdam, The Netherlands

**Abstract**

**High-energy (1-100 keV) electrons can coherently couple to plasmonic and dielectric nanostructures creating cathodoluminescence (CL) of which the spectral features reveal details of the material's resonant modes at deep-subwavelength spatial resolution. While CL provides fundamental insight in optical modes, detecting its phase has remained elusive. Here, we introduce Fourier-transform CL holography as a method to determine the far-field phase distribution of scattered plasmonic fields. We record far-field interferences between a transition radiation reference field and surface plasmons scattered from plasmonic nanoholes, nanocubes and helical nano-apertures and reconstruct the angle-resolved phase distributions. From the derived fields we derive the relative strength and phase of induced scattering dipoles. The data show that each electron wavepacket collapses at the sample surface and coherently excites transition radiation and surface plasmon quanta. Fourier-transform CL holography opens up a new world of coherent light scattering and surface wave studies with nanoscale spatial resolution.**

**Introduction**

Cathodoluminescence (CL) spectroscopy is a unique technique to create and probe optical materials excitations at nanoscale spatial resolution [1]. In coherent CL a sample is directly polarized by the time-varying electric fields carried by a high-energy (1-100 keV) electron beam [2,3]. Each electron creates a single electromagnetic field cycle in the sample with a duration of a few hundred attoseconds with a corresponding frequency spectrum with energies in the 0-30 eV spectral range. The electron thus acts as a broadband source of optical excitation with a spatial resolution limited by the extent of its evanescent field (~10 nm) [4]. The electron-induced polarization excitations can then decay by optical radiation (CL) that is collected in the far field.

CL spectroscopy directly probes the radiative local density of optical states, and spatial maps of the CL spectrum probe detailed information on optical modes in photonic nanostructures at deep-subwavelength spatial resolution [2,4–6]. Several CL modalities have been developed recently: angle-resolved CL spectroscopy provides a direct measure of photonic bandstructures [3,4]; CL polarimetry provides the full polarization state [7], and $g^{(2)}(\tau)$ two-photon correlation spectroscopy provides quantum statistics of emitted CL photons [8,9]. A key missing parameter in CL spectroscopy so far is measurement of the wavefront of the emitted light. Phase information is crucial to reconstruct the nature of the electron-induced polarization densities in CL spectroscopy, and in general, to control the structure of scattered optical wavefronts which is key to many applications in imaging, integrated optics, optical computing, optical communication and more.

Here, we introduce Fourier-transform CL holography as a method to determine the far-field phase distribution of scattered fields with nanoscale spatial excitation resolution [10–12]. Fourier-transform holography was previously applied in other fields [13–17]. In our work we analyze the CL signal that originates from electron-beam excited surface plasmon polaritons (SPPs), hybrid light-matter waves which propagate in two dimensions at the interface between a metal and a dielectric. Their strong electric and magnetic fields confined to the interface provide unique ways to control light-matter interactions at the nanoscale. SPPs can carry information in miniature integrated circuits [18], enable efficient sensors [19], and couple efficiently to quantum emitters [20,21]. In all these 2D-confined geometries precise control over SPP scattering is essential in order to control coupling to the third dimension.

We use 30 keV electrons to excite SPPs that propagate at a Ag/SiO$_x$ interface and subsequently scatter by suitably designed plasmonic scatterers. The scattered light interferes with transition radiation (TR) that is excited by the same electron at the point of impact and that serves as a reference field with known phase and polarization [3,22]. Applying Fourier-transform CL holography we reconstruct the angle-resolved amplitude and phase distribution of the p-polarized scattered fields originating from a subwavelength hole in a Ag film. Applying the same technique to a single-crystal Ag nanocube (NC) deposited on the Ag/SiO$_x$ stack, which forms a gap plasmon resonance, we retrieve the $\pi$ phase jump for light scattered at frequencies across the resonance. Moreover, Fourier-transform CL holography reveals that helical nano-apertures made in a single-crystal gold surface convert SPPs to free-space waves with pronounced phase singularities. From the data we conclude that each temporally coherent electron wavepacket collapses to a single point charge at the point of impact [23], simultaneously exciting transition radiation and surface plasmon polariton quanta. The measurements take advantage of the 10 nm spatial resolution of CL excitation spectroscopy, establishing Fourier-transform CL holography as a powerful deep-subwavelength technique to study scattering phenomena of surface waves and (resonant) nanostructures.

**Results**

**Plasmonic scattering geometries.** A 200-nm-thick Ag film was deposited on a Si(100) substrate using thermal evaporation. Subsequently a 15-nm-thin SiO$_x$ film was deposited to avoid oxidation of the Ag film. Spectroscopic ellipsometry was used to characterize the optical constants and layer thicknesses (see section 1 of SI). From these data we derived the SPP dispersion at the Ag/SiO$_x$ interface; the SPP mode effective index and the propagation length are shown in Fig. 1c. Focused ion-beam milling using 30-keV Ga ions was used to fabricate 300-nm-diameter holes in the Ag/SiO$_x$ stack with a depth of 215 nm [see Fig. 1b (top)]. On the same multilayer substrate we drop-casted 75-nm-sized single-crystalline Ag nanocubes (NCs) [see Fig. 1b (bottom)]. The NCs were made using the synthesis procedure as described in Ref. [24]. This procedure leads to a highly monodisperse solution of Ag NCs that are functionalized by polyvinylpyrrolidone (PVP) ligands. Fabrication details on thin-film growth and chemical synthesis of Ag nanocubes are given in the Methods section. Focused-ion beam milling using 30-keV Ga ions was used to fabricate helical nano-apertures on the (111) surface of polished single-crystalline Au. The helical nano-apertures have a diameter of 940 nm and a maximum depth of 570 nm.

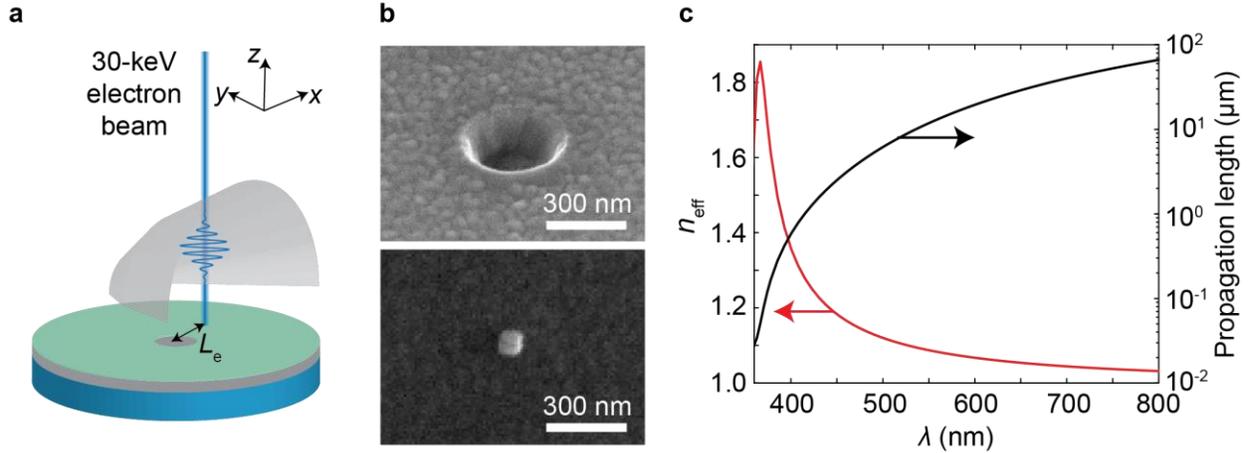

**Figure 1 | Cathodoluminescence geometry, Ag sample and surface plasmon polariton dispersion. (a)** Schematic of experiment. A 30-keV electron wave packet collapses onto a $SiO_x$ (15 nm)/Ag (200 nm)/Si stack and generates broadband transition radiation (TR) and surface plasmon polaritons (SPP). The SPPs scatter from a nanohole, nanocube or nanohelix that is placed at a distance $L_e$ from the excitation point. Both TR and radiation from scattered SPPs are collimated by a parabolic mirror and projected onto a CCD array. **(b)** SEM images of 215 nm deep 300-nm-diameter nanohole in the layer stack (top) and 75-nm Ag nanocube on the layer stack (bottom). **(c)** Effective index of SPP mode: Re($n_{SPP}$) and propagation length: $1/[2\text{Im}(k_0 n_{SPP})]$ of SPP mode derived from the optical constants obtained from spectroscopic ellipsometry of the layer stack.

**Cathodoluminescence spectroscopy.** CL experiments were performed using a scanning electron microscope (SEM, $V_{acc}$=30 kV, $I$=4 nA) equipped with an aluminum paraboloid mirror to collect CL. We use two measurement geometries: (1) angle-resolved, or Fourier, CL (ARCL) in which a wide zenithal and azimuthal angular emission pattern is collected over a wavelength range determined by a band pass filter ($\lambda$ = 600±20 nm) [25]; and (2) hyper-spectral angle-resolved cathodoluminescence (HSARCL) in which the angular distribution of CL intensity is collected in the vertical plane along the parabola's center and analyzed with a spectrometer at 0.9 nm spectral resolution [26,27]. Details of the CL geometry and spectroscopy are described in the Methods section.

**Scattering from plasmonic nanoholes in Ag.** Figure 2a shows the ARCL intensity at $\lambda$=600 nm for the electron beam placed $L_e$=2.29 μm away from the center of the nanohole. The electron beam is positioned on the right side of the nanohole, along the horizontal x axis crossing the center of the nanohole (see Fig. 1a). A clear interference pattern is observed with fringes along the vertical $k_y$ direction [22]. Figure 2b shows the ARCL intensity at $\lambda$=600 nm for the unstructured planar stack. A cylindrically-symmetric transition radiation pattern is observed, with the highest intensity observed at larger zenithal angles as expected for an upward-oriented dipole slightly above the multilayer stack that represents transition radiation. Subtracting the transition radiation reference from the data of Fig. 2a results in a pronounced interference pattern as shown in Fig. 2c. The modulation depth of the fringes is 45% in the forward scattering direction, and lower for backward scattering (see section 2 of SI on how we derived the visibility of the interference fringes and for a comparison with numerical simulations). Taking the 2D fast Fourier transform of the data presented in Fig. 2c using $k_0=2\pi/\lambda$ results in Fig. 2d (see section 3 of the SI for the convention used for the Fourier transform). Aside from a central spot two distinct spots are observed that are displaced 2.29 μm from the origin, in agreement with the distance between the electron beam and the center of the hole in the experiment. As shown in section 4 of the SI these spots contain the interference terms, while the central spot represents the scattered intensity coming from individual scattering centers.

After isolating the interference term on the left in Fig. 2d (see section 4 of SI), and performing a 2D inverse fast Fourier transform, we retrieve the angular amplitude and phase patterns of the p-polarized scattered electric field $E_{sc,p}$. The amplitude profile shown in Fig. 2e shows a bright leftward-oriented lobe and a weak rightward lobe along the horizontal $k_x$ axis. Figure 2f shows the phase pattern which shows a π phase flip between left and right propagating fields, all relative to the phase of the spherical transition radiation wave front which varies only very weakly (see section 5 of SI). This shows that the induced polarization density in the nanohole breaks cylindrical symmetry and has a strong in-plane component. This analysis now provides the full (p-polarized) electric field amplitude and phase of the field radiated by the SPP-driven nanohole.

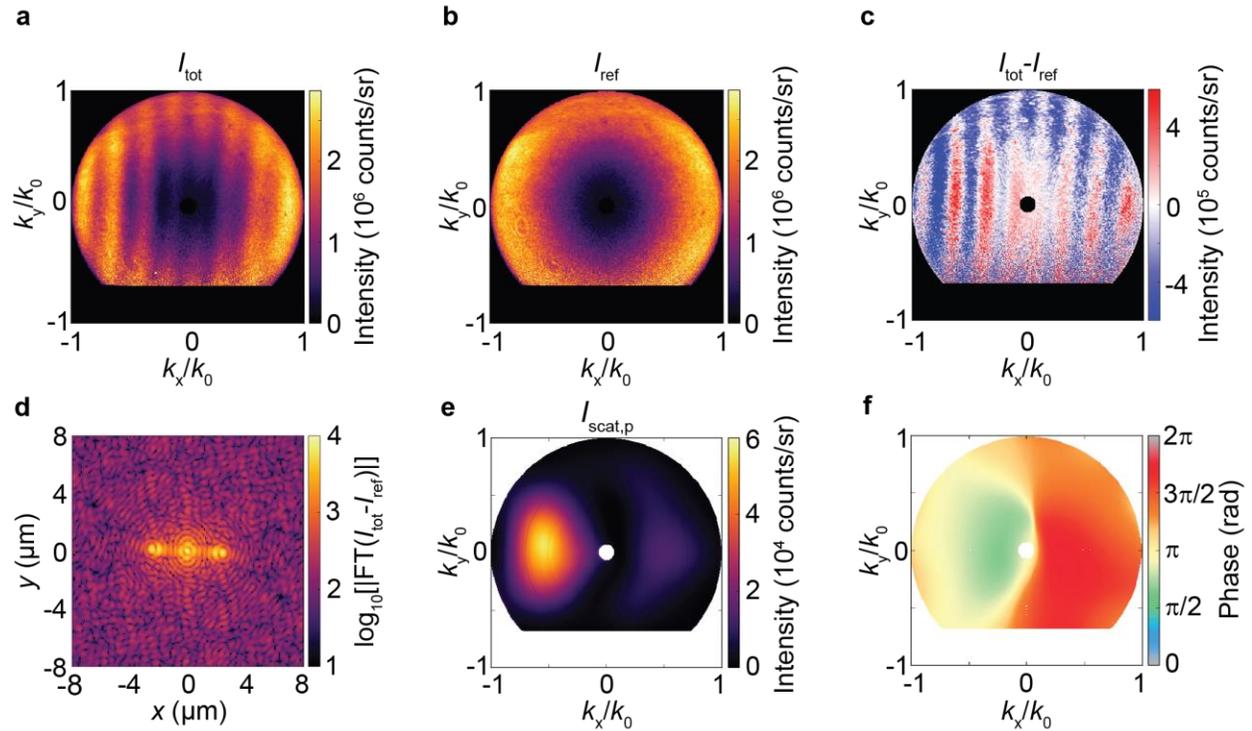

**Figure 2 | Phase profiles for SPPs scattered from a nanohole in Ag. (a)** Angle-resolved cathodoluminescence radiation pattern (λ= 600±20 nm) obtained for a 30-keV electron beam placed 2.29 μm to the right of a 300-nm-diameter nanohole. **(b)** Transition radiation from the same layer stack in the absence of nanoscatterer. **(c)** The difference of data presented in **(a)** and **(b)**. **(d)** 2D fast Fourier transform of data presented in **(c)**. **(e)** Numerically derived p-polarized amplitude pattern for the nanohole. **(f)** Numerically derived phase profile of p-polarized scattered field by the nanohole.

As we have shown previously, nanoholes in metal films possess both electric and magnetic dipoles that interfere in the far field resulting in strong angular beaming from the hole [28,29]. Using the amplitude and phase information from the analysis above, we can directly perform a multipolar decomposition of the scattered fields (for details on this calculation, see section 6 of SI). Given the small size of the hole we limit the multipolar decomposition to electric and magnetic dipoles and find that SPP excitation of the Ag nanohole induces mainly x- and z- oriented electric ($p_x$, $p_z$) and y-oriented magnetic ($m_y$) dipoles [28]. This is in full agreement with the fact that transverse-magnetic (TM)-polarized SPPs propagating along the x axis contain x and z electric field and y magnetic field components that directly couple to these three dipole moments. Figure 3a graphically presents the retrieved complex-valued dipole moments. We find that the phase difference $\phi$ between $m_y$ and $p_x$ is $0.5\pi$, in full agreement with the Maxwell-Faraday relation that states that the current loop formed by the $p_x$ dipole and its image dipole induces a $\pi/2$-phase-shifted magnetic field [$\nabla \times E(r,\omega) = i\omega\mu_0 H(r,\omega)$]. The fact that small $p_y$ and $m_x$ contributions are also found, despite the symmetry of the scattering problem, is ascribed to the fact that the parabolic mirror causes the far fields created by $p_z$, $p_y$, and $m_x$ to be non-orthogonal (see Table S3). The excitation of $p_z$ then leads to small apparent $p_y$ and $m_x$ components in the measurement. The asymmetric beaming of the scattered radiation (Fig. 2e) is a direct consequence of the interference between the induced z-polarized electric dipole and the induced in-plane dipoles [29].

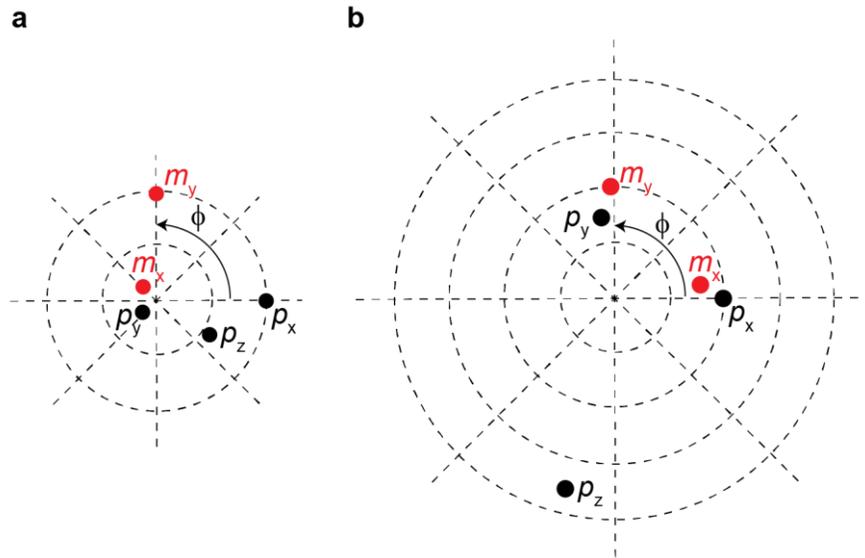

**Figure 3 | Multipole decomposition of the scattered field from CL holograpy data.** Visualization of numerically retrieved values of electric (black) and magnetic (red) dipole moments. The amplitude of the dipoles is encoded in the radial distance to the center of the circle; the phase in the polar angle $\phi$. **(a)** Electron beam is placed $L_e$=2.29 μm to the right of a nanohole. **(b)** Electron beam is placed $L_e$=2.29 μm to the right of the Ag nanocube.

Next, we performed HSARCL in which we measure the spectrally resolved angular radiation pattern in the plane defined by the surface normal and the $k_y$ axis in Fig. 2a. We rotated the excitation scheme presented in Fig. 1a by 90° clockwise, including the Cartesian coordinate system, so that the interference fringes can be measured. This measurement scheme allows probing fine details in the interference phenomena over a wide wavelength range, enabling derivation of the SPP dispersion as well as the phase distribution of the scatterer's emission profile, as we will show. Figure 4a shows the CL dispersion diagram in the $\lambda$=375-820 nm spectral band for the electron beam placed $L_e$=2.40 μm away from the center of the nanohole along the positive x axis. Subtracting the transition radiation reference data (Fig. 4b), we find clear interference fringes over a broad wavelength range (Fig. 4c). We model the dispersion relation of the maxima of the interference fringes by considering the constructive interference conditions in the far field taking into account the optical path lengths ($k_0 n_{SPP} L_e$, $k_x L_e$), with $k_x$ the in-plane wave vector of scattered light, and a phase term $\phi_{scat}$ related to scattering:

$$\lambda_m = \frac{L_e}{N - \frac{\phi_{scat}}{2\pi}} \left( \frac{k_x}{k_0} + n_{SPP} \right). \tag{1}$$

The solid curves in Fig. 4c are fits of this model to the data for different orders N; they fit the data very well over a broad spectral range, with N ranging from N=1-5, all for the same scattering phase ($\phi_{scat}=0.7\pi$). The curvature of the solid curves clearly reflects the SPP dispersion; the dashed lines show the dispersionless case ($n_{spp}=1$) for reference. The largest deviation between the solid and dashed curves occurs for shorter wavelengths, in agreement with the SPP dispersion shown in Fig. 1c. The data in Fig. 4a-c show that the scattering of SPPs by the nanohole is a non-dispersive process in this wavelength range described by a fixed overall scattering phase of $0.7\pi$.

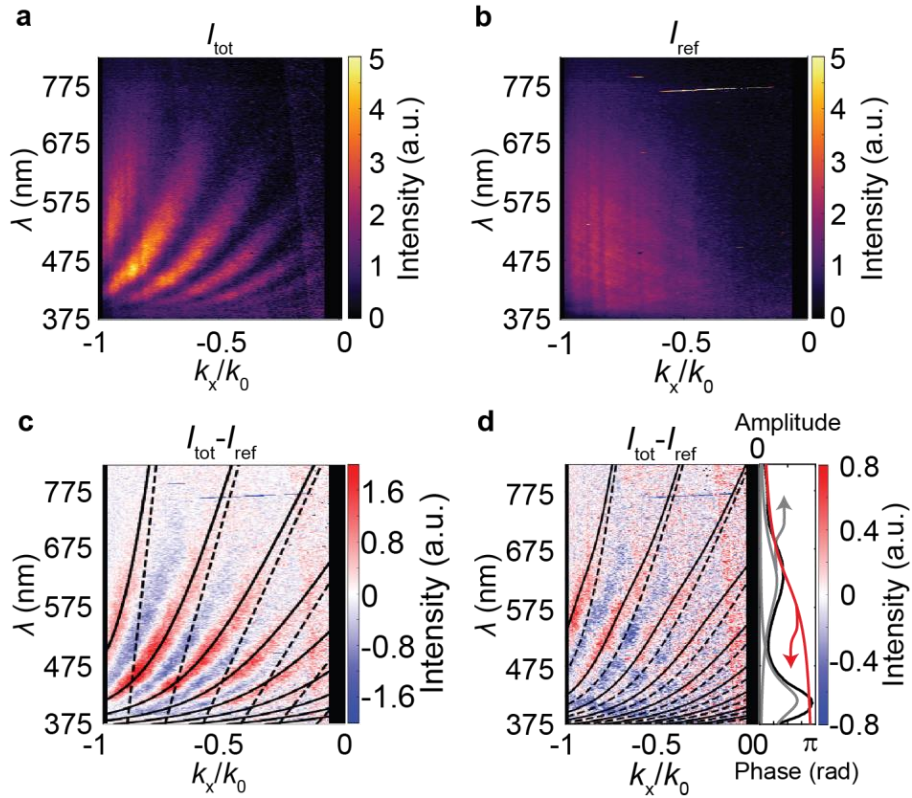

**Figure 4 | Plasmon dispersion and resonant scattering revealed with hyper-spectral angle-resolved cathodoluminescence. (a)** CL radiation pattern for electron beam placed $L_e$=2.40 µm to the right of the center of a nanohole ($I_{tot}$). **(b)** Transition radiation pattern for the stack ($I_{ref}$). The vertical black sections in **(a)** and **(b)** reflect the hole in the parabolic mirror through which the electron beam enters. **(c)** Difference of data presented in **(a)** and **(b)**. Solid curve is fit to constructive interference wavelength $\lambda_m$ given in Eq. 1. Dashed curve assumes a dispersionless surface wave. **(d)** Left: $I_{tot} - I_{ref}$ for electron

beam placed $L_e$=2.37 μm to the right of a Ag nanocube. Solid curve is fit to constructive interference condition, taking into account a Lorentzian scattering resonance. Dashed curve includes a dispersive surface wave, and does not include the resonance. Right: Lorentzian resonance considered for fit (black curve: measured CL spectrum, red curve: fitted phase, grey curve: fitted amplitude).

**Scattering from Ag nanocubes.** Next, we investigate the scattering of SPPs by single-crystalline Ag NCs [30]. It is known that NCs on a metal substrate with a thin dielectric spacer possess gap plasmon resonances in the visible spectral range [31]. We collect ARCL data at $\lambda$=600 nm using the same distance between electron impact position and the center of the NC of $L_e$=2.29 μm. Subtracting the transition radiation, we find the angular profile in Fig. 5a. Clear interference fringes are observed, with a notable left/right asymmetry. To study this in more detail we plot in Fig. 5b the angular data averaged along the vertical angular axis for scattering from the hole and the NC. In both cases, the fringe amplitude is higher for negative than for positive $k_x$ values which we relate to the left/right symmetry breaking by the off-center excitation in combination with the use of a bandpass filter (see SI). Moreover, for the Ag NC the maxima for positive $k_x$ values are phase-shifted compared to values for negative $k_x$, while this phase shift is not observed for the hole. This reflects a fundamental difference in the SPPs scattering mechanism for NCs and nanoholes, as we will further illustrate below.

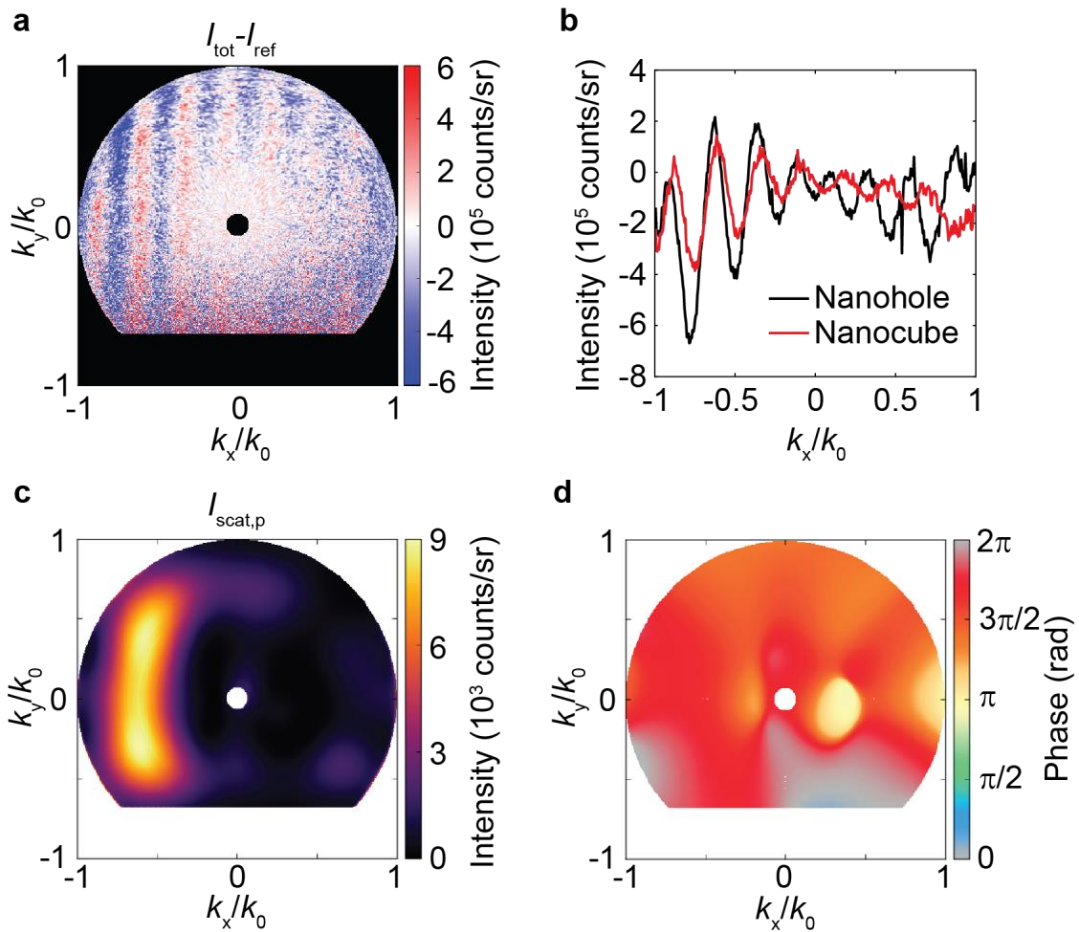

**Figure 5 | Angular intensity and phase profiles for SPPs scattered from Ag nanocube. (a)** Angle-resolved cathodoluminescence radiation pattern ($\lambda$=600±20 nm) corrected for transition radiation. **(b)**

Interference fringes obtained by averaging data of Fig. 2c and Fig. 5a along the $k_y/k_0$ axis. **(c)** Numerically derived p-polarized amplitude pattern for the nanocube. **(d)** Numerically derived phase profile of p-polarized scattered field by the nanocube.

Using Fourier analysis we retrieve both the amplitude (Fig. 5c) and phase (Fig. 5d) patterns of the p-polarized scattered electric field radiated by the Ag NC. As for the nanohole, we observe beaming of light to the left, which we partially attribute to the asymmetry in the excitation process and the resulting modal excitation. In contrast to the case of the hole, the phase profile is found to be quite homogeneous in angle, i.e., the $\pi$ phase flip between left and right propagating fields as was observed for scattering off the hole, is not observed here. Using the intensity and phase patterns we perform a multipolar decomposition of the scattered fields (see section 6 of SI) and find that the main scattering contributions come from $p_x$, $p_y$, $p_z$, $m_x$ and $m_y$ (see Fig. 3b for the graphical representation of the retrieved complex-valued dipole moments). We find similar amplitudes for $p_x$ and $m_y$ and for $p_y$ and $m_x$ with a phase difference between electric and magnetic dipoles of $0.5\pi$ and $-0.5\pi$, respectively, again explained by the electric-magnetic dipole coupling argument described above for the nanohole. The nanoparticle shows a dominant z-polarized electric dipole mode, as expected for a plasmonic nanocube above a mirror with a dielectric spacer in between [32]. By symmetry, the angular profile of the phase is expected to be symmetric, as is observed. The contrasting phase symmetries observed for hole and nanocube are clearly reflected in the shifted phase profiles for positive angles in Fig. 5b.

Next, we present the HSARCL data for the Ag NC obtained by placing the electron beam at $L_e$=2.37 μm in Fig. 4d. As for the nanoholes we observe a clear interference pattern over a broad spectral range, with mode numbers N=1-5. However, for the case of the NCs, the fringes do not match the dispersive SPP model indicated by the dashed curves. The discrepancy is a direct manifestation of the increasing phase shift between the induced polarization density and the plasmonic driving field as the wavelength is decreased across the scattering resonance. Figure 4d shows the CL spectrum of the Ag NC taken by directly placing the electron beam at a corner of the NC. As can be seen, the plasmon spectrum peaks at 620 nm with a full-width-at-half-maximum linewidth of 152 nm. Indeed, the largest discrepancy between the data in Fig. 4d and the dispersive SPP model occurs for wavelengths below the resonance peak. To quantitatively analyze this trend, we fit the plasmon contribution to the CL spectrum with a single Lorentzian line shape and introduce the corresponding phase shift to the dispersion model (solid curve). Clearly, this resonant scattering model fits the trends in the CL data well for different values of N. This analysis clearly shows the power and sensitivity of CL holography to detect characteristic phase shifts in scattering.

**Observation of phase singularity.** As a final demonstration of the power of CL holography we study SPP scattering from a helical nano-aperture in single-crystalline Au (Fig. 6a). Figure 6b shows the ARCL intensity at $\lambda$=600 nm for the electron beam placed $L_e$=2.29 μm away from the center of the nanohelix along the negative y axis. A clear interference pattern is observed with fringes along the $k_y$ direction. Remarkably, we observe a clear fork-like structure around $(k_x/k_0, k_y/k_0)$=(0.25;0.60) pointing at the existence of a phase singularity in the far-field phase profile. The retrieved angle-dependent intensity and phase patterns of the scattered field are shown in Fig. 6c, and 6d respectively. Figure 6c shows the radiation profile is strongly beamed in the forward direction. Figure 6d shows for $(k_x/k_0, k_y/k_0)$=(0.28;0.60) a phase singularity with topological charge -1, as the phase evolves once from 0 to $2\pi$ when turning clockwise around the phase singularity.

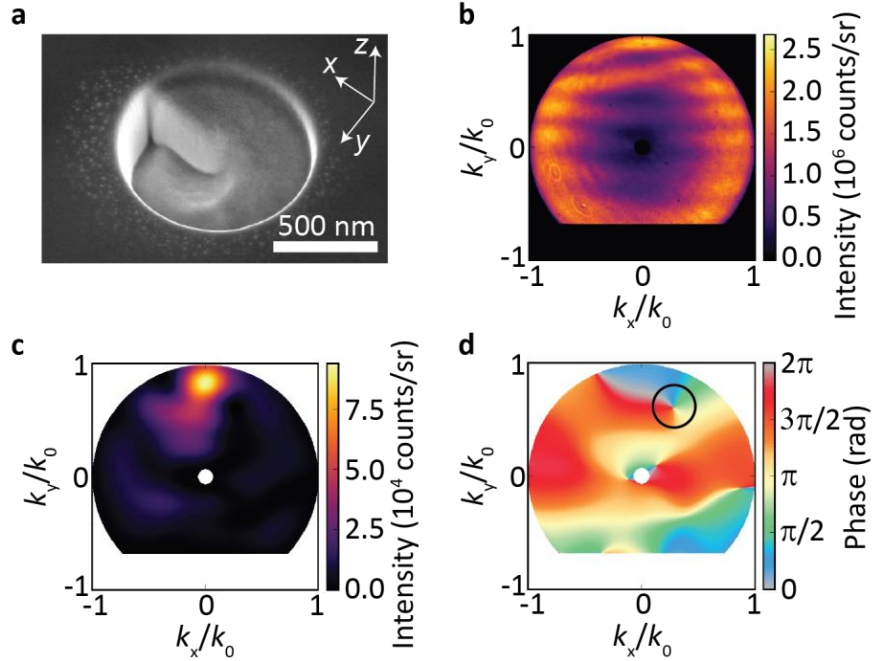

**Figure 6 | Phase singularity for SPPs scattered from a helical nano-aperture in Au. (a)** SEM image of helical nano-aperture milled in single-crystalline Au. **(b)** Angle-resolved cathodoluminescence radiation pattern ($\lambda$= 600±20 nm) obtained for a 30-keV electron beam placed 2.29 µm from the helical nano-aperture along the negative y axis. **(c)** Numerically derived p-polarized intensity pattern for the helical nano-aperture. **(d)** Numerically derived phase profile of p-polarized scattered field by the helical nano-aperture (indicated by circle) showing phase singularity with topological charge -1.

**Electron wavepacket collapse.** The CL holography data provide a unique opportunity to test fundamental models of coherent electron excitation of matter, where the electron possesses both wave and particle properties. The interference analysis used in the three complementary holography experiments described above assumes the electron acts as a particle with a point charge, where each electron simultaneously excites a TR and SPP quantum of which the interference is set by the moment of electron impact (interaction between multiple electron impacts can be neglected at the used beam current). However, it is well known that individual electrons also act as wavepackets with a temporal and spatial coherence determined by the electron source. Given the energy spread of the source in our SEM the electron wavepacket as it hits the sample is about 6 µm in length. This corresponds to many optical cycles in the generated light so one could wonder if the particle nature is applicable and if this temporal incoherence would smear out the interference of the coherently excited TR and plasmon quanta [33,34].

The time-varying electric fields that polarize the sample in CL result from the time-varying current density of the moving electron which is given by the Dirac equation $j = \Psi^H \bar{\bar{\alpha}} \Psi$, with $\Psi(x,y,z,t)$ the wavefunction of the electron and $\bar{\bar{\alpha}} = \langle \overline{\alpha_1}, \overline{\alpha_2}, \overline{\alpha_3} \rangle$ the Dirac matrix. In the recoil-less limit the final wavefunction is the same as the initial wavefunction, so that $j \in \mathbb{R}^3$, which implies the current density does not have a phase that continuously varies in time and space. Therefore the shape and extent of the electronic wavepacket is irrelevant in the excitation mechanism. The same conclusion also follows from the first-order time-dependent perturbation theory presented in Ref. [33] from which it was concluded that spontaneous CL processes are independent of electron wavepacket size. This leads to an interpretation that CL is the incoherent sum of far-field radiation originating from electronic wavepackets for which the wavefunctions collapsed at a specific spot and specific time. We therefore conclude that the electron wavepacket collapses in time as it interacts with the sample, effectively acting as a point charge generating the TR and SPP quanta. This result is complementary to Ref. [23] which investigated the effect of spatial coherence of the electron wavepacket as it generates Smith-Purcell radiation from a nanoscale grating. There it was found that the electron wavepacket collapses in the spatial domain as it creates radiation [23]. Taking these data together, we conclude that electron wavepackets collapse in both time and space as they polarize matter, a result that is relevant for a broad range of coherent CL and electron energy loss spectroscopy (EELS) experiments.

To investigate if the point charge excitation model quantitatively explains the visibility of the observed fringes we performed finite-difference time domain (FDTD) simulations of the plasmon scattering and far-field interference processes and calculated the visibility of the interference fringes for the nanocube (see section 2 of SI). We find that within uncertainties of the simulations due to uncertainties in fine details of the scattering geometries the experimentally observed fringe visibilities agree with the simulations, supporting the collapsing electron wavefunction model.

**Conclusion**

We have introduced Fourier-transform cathodoluminescence holography as a technique to directly retrieve the phase of light scattered from nanoscale objects. The incident electrons generate surface plasmon polaritons on silver that effectively scatter from nanoscale holes, nanocubes, and nanohelices, and at the same time induce a transition radiation reference field. Far-field interference between scattered SPPs and the broadband transition radiation allows retrieval of both intensity and phase of the scattered fields. From the data, we retrieve the phase and amplitude of the 3D vectorial electric and magnetic dipole moments that interfere to create the measured scattering pattern. The spectral phase jump across a plasmonic scattering resonance is directly revealed from the data and we directly observe phase singularities in the scattering of plasmonic nanohelices. Fourier-transform cathodoluminescence holography can find many further applications to probe scattering of surface waves and other guided waves from nanoscale objects at high precision. The data show that each temporally dispersed electron wavepacket collapses at the sample surface to create coherently coupled transition radiation and surface plasmon polariton quanta.

**Methods**

**Fabrication of multilayer stack.** A layer stack of 200 nm Ag and 15 nm $SiO_x$ was made by thermal evaporation. Ag was deposited at a base pressure of $1.6×10^{-6}$ mbar and a deposition rate of 2 Å/s. For the $SiO_x$ layer we used $SiO_x$ as the target; a deposition rate of 0.6 Å/s and a base pressure of $9.3×10^{-7}$ mbar.

**Synthesis of Ag nanocubes.** The Ag nanocubes were synthesized by adopting a chemical synthesis procedure reported earlier [35]. Well-defined (100)-faceted Ag cubes of ~75 nm were made in solution, filtrated and dispersed in ethanol, and then dropcast onto the multilayer stack.

**Fabrication of nanohelices.** 30 keV Ga focused-ion beam milling of nanohelices was performed using serpentine scans at a current of 1.5 pA and a pixel dwell time of 1.5 μs. The structure was made in 60 passes, each pass taking 667 ms.

**CL measurement geometries.** CL experiments were performed using a Thermo Fisher 650 Quanta SEM equipped with a thermionic Schottky field emission electron source, operated at 30 kV and a typical beam current on the sample of 4 nA. The electron energy spread after the source is 0.7 eV, and the drift distance from acceleration section to sample is 50 cm. The CL is collected with an aluminum paraboloid mirror ($1.47\pi$ sr acceptance angle), of which the focus is aligned with respect to the electron beam and the sample using a motorized micropositioning stage inside the vacuum chamber. The CL signal is analyzed using a Delmic SPARC system equipped with a 2048×512 pixel back-illuminated CCD array mounted on a Czerny-Turner spectrograph. We perform two different types of CL experiments:

(1) Angle-resolved CL (ARCL): in this geometry, the spectrometer slit is fully opened (15 mm) and a planar aluminum mirror is selected on the turret in the spectrograph. The reflected light is projected onto the CCD camera. In this way, we acquire a full angular pattern/momentum distribution within the NA of the paraboloid collection optics. Wavelength specificity is attained with a band-pass filter ($\lambda$=600±20 nm).

(2) Hyper-spectral angle-resolved cathodoluminescence (HSARCL): the entrance slit is closed to 150 μm and acts as a filter in angular/momentum space, selecting the radiation in the vertical plane along the paraboloid optical axis. This angular emission pattern is dispersed by the diffraction grating in the spectrometer, leading to a hybrid 2D CCD image with wavelength on the horizontal axis and angle on the vertical axis. This map is converted into a wavelength ($\lambda$) and momentum ($k$) map by applying the appropriate coordinate transform.

**Acknowledgements**

We gratefully acknowledge Bob Drent for help with the fabrication of the multilayer stack, Dimitri Lamers for FIB nanofabrication, and Toon Coenen for insightful discussions. This work is part of the research program of AMOLF which is partly financed by the Dutch Research Council (NWO). This project has received funding from the European Research Council (ERC) under the European Union's Horizon 2020 research and innovation programme (grant agreement No. 695343) and an NWO VIDI grant (project number 14846).

**Author contributions**

N.J.S. and A.P. conceived the experiment. N.J.S. performed the CL experiments, data analysis, and theoretical calculations. H.A. fabricated the Ag nanocubes. All authors contributed to the manuscript. E.C.G. and A.P. gave overall supervision.

**Additional information**

**Competing financial interests:** A.P. is co-founder and co-owner of Delmic BV, a company that produces the cathodoluminescence system that was used in this work.

# SUPPLEMENTARY INFORMATION

## 1. Optical constants of the stack

Figure S1 shows the dispersion for both Ag (Fig. S1a) and SiO$_x$ (Fig. S1b) retrieved from spectroscopic ellipsometry.

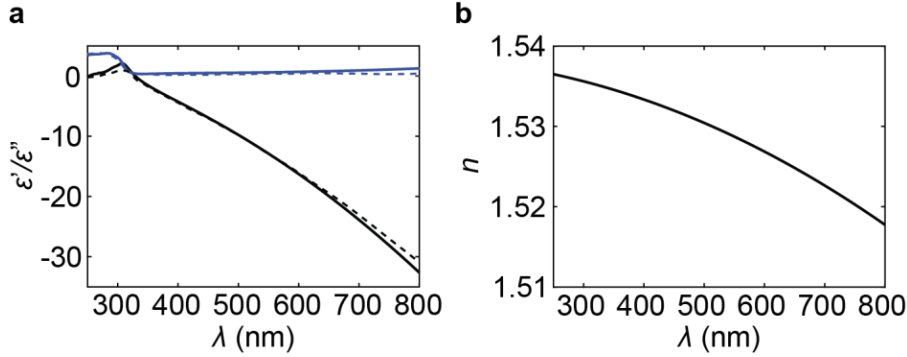

**Figure S1 | Optical constants derived from spectroscopic ellipsometry data. (a)** Dielectric constant of evaporated Ag (solid curves) and from Johnson and Christy (dashed curves). The real/imaginary part of the dielectric constants are shown in black/blue. **(b)** Refractive index of evaporated SiO$_x$.

## 2. Angle-dependent visibility for surface plasmon scattering from nanocube

Since the angular scattering intensities for transition radiation and surface plasmon scattering are not the same, the interference fringe visibility is angle-dependent. From the retrieved scattered intensities of SPPs and TR (Fig. 5) we calculated the angle-dependent fringe visibility $V$ for the nanocube

$$V = \frac{2\sqrt{I_{TR}I_{sc,p}}}{I_{TR}+I_{sc,p}}. \qquad (S1)$$

over the entire azimuthal and zenithal angle range (Fig. S2a). We find that for the angular range along the $k_x$ axis, where the CL intensity is highest ($k_x/k_0$=-0.7--0.6) and the data most accurate, $V$=0.15.

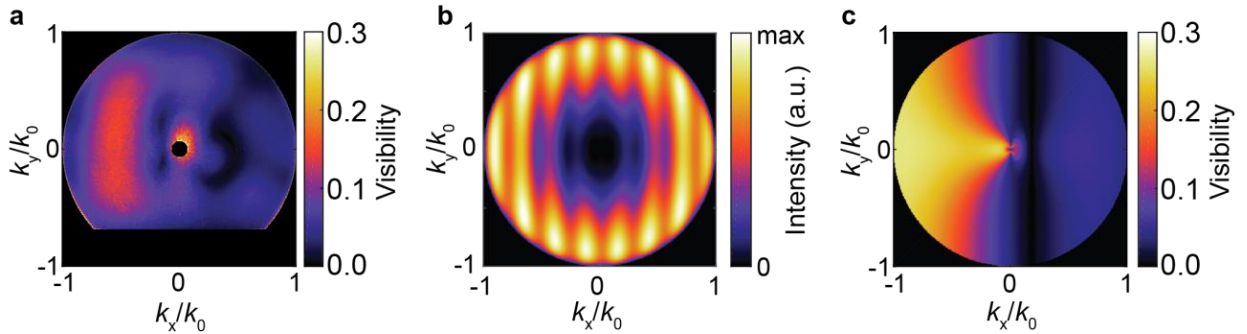

**Figure S2 | Visibility of far-field interference fringes for the nanocube. (a)** Angle-dependent fringe visibility derived from Fourier analysis data in Fig. 5. **(b)** Simulated far-field angular emission pattern for a z-polarized electric dipole ($\lambda$=600 nm) placed 10 nm above the stack and 2.29 μm to the right of

a Ag nanocube. **(c)** Angle-dependent visibility as calculated according to Eq. S1, multiplied by the sinc function that mimics smearing out of interference fringes due to finite bandwidth.

We calculated the angular emission pattern using finite-difference time domain (FDTD) simulations with the electron impact simulated by a z-polarized electric dipole placed 10 nm above the stack and 2.29 μm to the right of a Ag nanocube. We use nanocube dimensions obtained from SEM images and optical constants derived from ellipsometry. Figure S2b shows the numerically calculated angular emission pattern, clearly showing the interference fringes. Using Eq. S1, we calculate the angular visibility of the fringes and find $V$=0.31 for $k_x/k_0$=-0.6. In experiments the visibility of the interference fringes is lowered due to the use of a $\Delta\lambda$=40 nm bandwidth color filter. To correct for this we multiply the interference term in the numerator of Eqn. (S1) by $|\text{sinc}\{\frac{\Delta\lambda}{2\lambda_c}k_0L_e[n_{\text{eff}}(\lambda_c)+\sin\theta]\}|$ and find the fringe visibility for the cube is reduced to 0.25 for $k_x/k_0$=-0.6 and to 0.06 for $k_x/k_0$=0.6 due to the use of the bandpass filter (see Fig. S2c). The asymmetry of this correction for positive and negative angles largely explains the strong asymmetry in $V$ found from the experiment (Fig. S2a). Given the uncertainties of the simulations due to uncertainties in fine details of the scattering geometries we conclude that within error bars this is in agreement with the experimentally observed fringe visibility

## 3. Definition of Fourier transform

The convention of the Fourier transform is in this work is:

$$f(x,y) = \iint f(k_x,k_y)e^{i(k_x x+k_y y)}\frac{\mathrm{d}k_x \mathrm{d}k_y}{(2\pi)^2}, \quad \text{(S2a)}$$

$$f(k_x,k_y) = \iint f(x,y)e^{-i(k_x x+k_y y)}\mathrm{d}x\mathrm{d}y. \quad \text{(S2b)}$$

## 4. Numerical algorithm to retrieve both amplitude and phase of the scattered field

The numerical algorithm applied to retrieve both amplitude and phase of the scattered field is similar to the theoretical framework in Ref. [12]. The far-field electric field consists of the p-polarized electric field of transition radiation ($E_{\text{ref,p}}$) and the scattered electric field from the nanoscatterer that in general can have both an s- and p-polarized component ($E_{\text{sc,s}}$ and $E_{\text{sc,p}}$). The far-field radiation pattern can then readily be obtained:

$$I_{\text{tot}} = I_{\text{ref,p}} + I_{\text{sc,p}} + I_{\text{sc,s}} + E_{\text{ref,p}}E_{\text{sc,p}}^* + E_{\text{ref,p}}^*E_{\text{sc,p}}, \quad \text{(S3)}$$

where all quantities in Eq. S3 have two variables: $k_x/k_0$ and $k_y/k_0$ ($k_0$=2π/$\lambda_c$, with $\lambda_c$ is the central wavelength of the band-pass filter). The in-plane position vector of the nanoscatterer $\boldsymbol{r}_{\text{sc}}$=⟨$x,y$⟩ is defined in a reference frame with the excitation point of the electron beam at the origin. Within the far-field approximation, the scattered electric field is given by

$$E_{\text{sc,p/s}} = E_{\text{sc,p/s}}^{(0)}e^{-i\langle k_x,k_y\rangle \cdot \boldsymbol{r}_{\text{sc}}}, \quad \text{(S4)}$$

with the superscript 0 indicating that the corresponding field has as origin $\boldsymbol{r}_{\text{sc}}$. Taking the 2D Fourier transform of the far-field intensity pattern results in

$$\tilde{I}_{\text{tot}}(\boldsymbol{r}) = \tilde{I}_{\text{ref,p}}(\boldsymbol{r}) + \tilde{I}_{\text{sc,p}}(\boldsymbol{r}) + \tilde{I}_{\text{sc,s}}(\boldsymbol{r}) + \tilde{E}_{\text{ref,p}}(\boldsymbol{r}+\boldsymbol{r}_{\text{sc}}) * \tilde{E}_{\text{sc,p}}^{(0)*}(\boldsymbol{r}+\boldsymbol{r}_{\text{sc}}) + \tilde{E}_{\text{ref,p}}^*(\boldsymbol{r}-\boldsymbol{r}_{\text{sc}}) * \tilde{E}_{\text{sc,p}}^{(0)}(\boldsymbol{r}-\boldsymbol{r}_{\text{sc}}).$$

(S5)

Both the scattered field and transition radiation originate from a subwavelength region, hence the spatial extent of all $\tilde{E}$ is roughly $\lambda$ (diffraction limit). The convolution of two functions that have both a spatial extent of $\lambda$ results in a function with a typical length scale $\sqrt{2}\lambda$. This means that the interference terms can be separated from the non-interference terms when the nanoscatterer is placed at least $\sqrt{2}\lambda$ away from the electron beam, as is the case here. We isolate the term $\tilde{E}_{\mathrm{ref,p}}(r+r_{\mathrm{sc}}) * \tilde{E}_{\mathrm{sc,p}}^{(0)*}(r+r_{\mathrm{sc}})$ by multiplying Eq. S5 with $\mathrm{rect}\left(\frac{|r-r_{\mathrm{sc}}|}{L}\right)$, where $L$=2 μm, and shift the data by $+r_{\mathrm{sc}}$, resulting in the interference term

$$\tilde{I}_{\mathrm{interference}}(r) = \tilde{E}_{\mathrm{ref,p}}(r) * \tilde{E}_{\mathrm{sc,p}}^{(0)*}(r)\mathrm{rect}\left(\frac{|r|}{L}\right). \quad (S6)$$

Performing the 2D inverse Fourier transform then results in an approximation of

$$I_{\mathrm{interference}} = E_{\mathrm{ref,p}} \cdot E_{\mathrm{sc,p}}^{(0)*}. \quad (S7)$$

Finally, the complex p-polarized scattered field can be obtained by:

$$E_{\mathrm{sc,p}}^{(0)} = \left(\frac{I_{\mathrm{interference}}}{\sqrt{I_{\mathrm{ref,p}}}}\right)^*. \quad (S8)$$

## 5. Far-field phase profile for transition radiation

Here, we show that the wave front for transition radiation in the SiO$_x$/Ag/Si stack is close to spherical and therefore uniform in our representation in ($k_x/k_0$, $k_y/k_0$) space. We calculate the far-field phase profile numerically by approximating the source by a z-polarized electric dipole 10 nm above the stack. Figure S4a shows from which angle $k_\parallel/k_0=\sin(\theta)$, the deviation of the phase is more than $\pi/10$ as compared to the phase for $k_\parallel/k_0$=0.001. From Fig. S4a we find that the phase profile of our reference field can be considered uniform when $k_\parallel/k_0$<0.9.

Next, we show in Fig. S4b the phase of transition radiation at $k_\parallel/k_0$=0.7 for $\lambda$=375-820 nm. The wavelength-dependent phase originates from the material dispersion of the stack. The phase difference within the wavelength range $\lambda$=580-620 nm is approximately $\pi/50$, which is negligible for the experimental results shown in Figs. 2, 5, and S2.

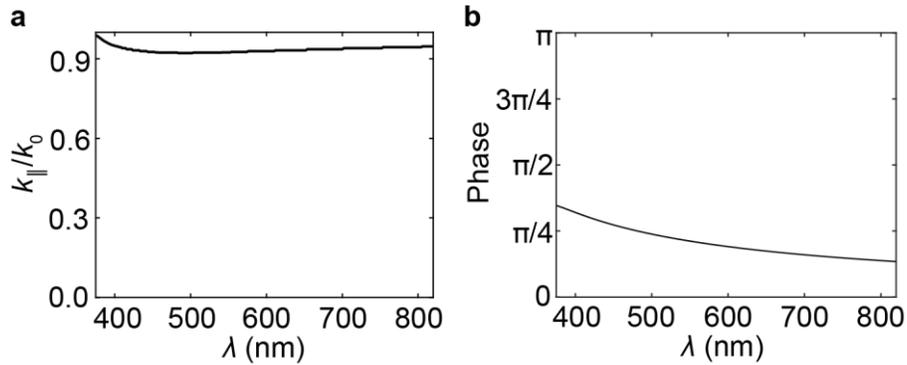

**Figure S4 | Far-field phase profile for transition radiation. (a)** Lower threshold value of $k_\parallel/k_0$ for which the transition radiation phase deviates more than $\pi/10$ from the phase at $k_\parallel/k_0$=0.001. **(b)** Far-field phase of transition radiation for $\lambda$=375-820 nm at $k_\parallel/k_0$=0.7.

## 6. Multipole expansion

In homogeneous space, the electric fields on a sphere can be decomposed in the orthonormal basis of vector spherical harmonics. Each basis function can be identified with one multipole radiating at the origin of the sphere. In this manuscript, we retrieve the (complex) p-polarized electric field in part of the upper hemisphere, defined by the parabolic mirror collecting CL. The retrieved p-polarized electric field in the upper hemisphere is decomposed in the non-orthogonal basis spanned by the p-polarized electric fields produced by multipoles placed at 10 nm above the stack. We limit ourselves to electric and magnetic dipoles. The p-polarized emission collected by the parabolic mirror forms 56%, 52% and 100% of the total emission that is collected for x-, y-, and z-polarized electric dipoles, respectively. For x-, y-, and z-polarized magnetic dipoles, this is 56%, 61%, and 0%, respectively. As we only retrieve the p-polarized electric field, and a z-polarized magnetic dipole uniquely emits s-polarized electric fields, we are insensitive to the z-polarized magnetic dipole. Hence, any emission originating from a z-polarized magnetic dipole cannot be retrieved with this experimental technique. We therefore omit this dipole in the analysis. All fields, including the retrieved electric fields, are normalized as follows:

$$\iint_{NA} |E_{MP,p}(k_x, k_y)|^2 dk_x dk_y = 1, \quad (S9)$$

where the NA is determined by the parabolic mirror. The amplitude and phase of the complex expansion coefficients are obtained as follows:

$$c_{MP} = \iint_{NA} E_{sc,p}(k_x, k_y) E^*_{MP,p}(k_x, k_y) dk_x dk_y. \quad (S10)$$

The coefficients for the nanohole and the nanocubes are listed in Table S1 and S2, respectively. A graphical representation of the results presented in Table S1 and S2 is presented in Fig. 3 of the main text. We found that the contribution of the different scattering components is somewhat sensitive to the precise shape of the hole: comparing measurements on different holes with slightly varying dimensions we find variations in amplitude of typically 10-20%, and with some exceptional larger deviations.

| Table S1 \| Coefficients of multipole expansion for SPP scattering from nanohole. | | | | | |
|---|---|---|---|---|---|
|  | $p_x$ | $p_y$ | $p_z$ | $m_x$ | $m_y$ |
| **Amplitude** | 0.80 | 0.13 | 0.46 | 0.14 | 0.78 |
| **Phase ($\pi$)** | 0.00 | 1.22 | 1.82 | 0.74 | 0.50 |
| The electron beam was placed $L_e$=2.29 $\mu$m to the right. | | | | | |

| Table S2 \| Coefficients of multipole expansion for SPP scattering from Ag nanocube. | | | | | |
|---|---|---|---|---|---|
|  | $p_x$ | $p_y$ | $p_z$ | $m_x$ | $m_y$ |
| **Amplitude** | 0.40 | 0.30 | 0.72 | 0.32 | 0.41 |
| **Phase ($\pi$)** | 0.00 | 0.55 | 1.42 | 0.05 | 0.51 |
| The electron beam was placed $L_e$=2.29 $\mu$m to the right. | | | | | |

Next, we check the orthogonality of the electric fields emitted by different dipoles(MP$_i$, and MP$_j$), by calculating

$$c_{MPij} = \iint_{NA} E_{MPi,p}(k_x, k_y) E^*_{MPj,p}(k_x, k_y) dk_x dk_y. \quad (S11)$$

The amplitude of $c_{MPij}$ is presented in Table S3. We observe that the electric fields emitted by different dipoles are not orthogonal. The non-zero value obtained for $p_z$ and $p_y$ can be explained by the limited collection angles along the y axis due to the parabolic mirror. As a consequence, we have observed in Fig. 3b of the main text that a strong z-polarized electric dipole comes together with a y-polarized electric dipole, even though this dipole cannot be excited due to symmetry arguments of the sample and excitation process. Even though the basis is not orthogonal and Eq. S10 is therefore strictly not valid, interesting trends can be learned from this analysis (see main text). For completeness, we provide an orthogonal basis for the in-plane dipoles that follows from a Gram-Schmidt process: $p_x+im_y$, $p_y+im_x$, $(p_x-im_y)-0.1i(p_x+im_y)$, $(p_y-im_x)+0.1i(p_y+im_x)$.

| | Table S3 \| Orthogonality of far-fields from all combinations of dipoles. | | | | |
|---|---|---|---|---|---|
| | $p_x$ | $p_y$ | $p_z$ | $m_x$ | $m_y$ |
| $p_x$ | 1.00 | 0.00 | 0.00 | 0.00 | 1.00 |
| $p_y$ | 0.00 | 1.00 | 0.14 | 1.00 | 0.00 |
| $p_z$ | 0.00 | 0.14 | 1.00 | 0.16 | 0.00 |
| $m_x$ | 0.00 | 1.00 | 0.16 | 1.00 | 0.00 |
| $m_y$ | 1.00 | 0.00 | 0.00 | 0.00 | 1.00 |
| Absolute value of overlap integral of far-field electric fields for all combinations of dipoles according to Eq. S10. | | | | | |

## 7. Surface plasmon polariton scattering from nanocubes

Figures S5a-c show the total (a) and reference (b) intensity from which data in Fig. 5 was derived, and the 2D fast Fourier transform of the difference (c). Figures S5d-f show the total intensity (d) for a NC where the electron beam was placed $L_e$=3.85 μm to the right of the NC, (e) the 2D fast Fourier transform of $I_{tot} - I_{ref}$ and (f) shows the retrieved far-field scattering pattern.

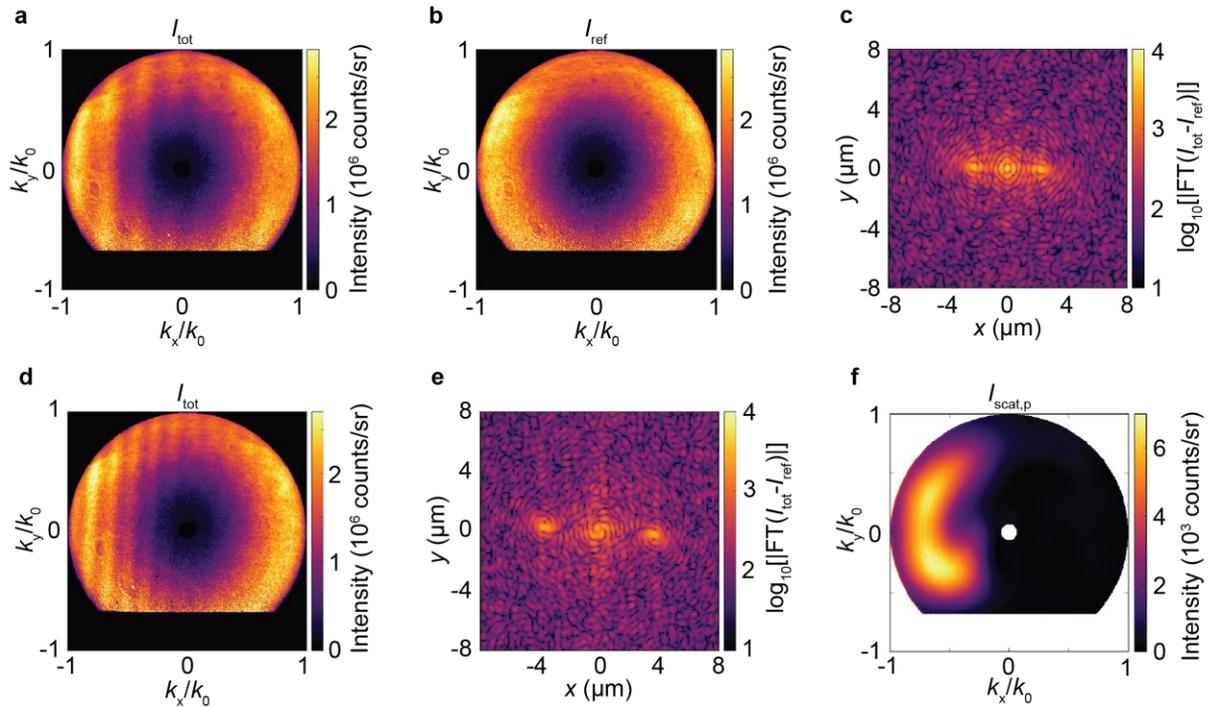

**Figure S5 | Interference fringes for surface plasmon polariton scattering by nanocube.** Experimental results for electron beam placed **(a-c)**: $L_e$=2.29 μm, **(d-f)**: $L_e$=3.85 μm to the right of a 75-nm Ag nanocube. **(a)** Angle-resolved cathodoluminescence radiation pattern filtered with a band pass color

filter ($\lambda$=600±20 nm) . **(b)** Reference measurement on same stack in absence of nanoscatterer. **(c)** 2D fast Fourier transform of the difference in intensity from Figs. S5**a-b**.